# 4D-Printing of Smart, Nacre-Inspired, Organic-Ceramic Composites

Benedikt F. Winhard[a,e], Philipp Haida[b], Alexander Plunkett[a], Julian Katz[a], Berta Domènech[a,c], Volker Abetz[b,d], Kaline P. Furlan[a,e], Gerold A. Schneider[a]

a Hamburg University of Technology, Institute of Advanced Ceramics, Denickestraße 15, 21073 Hamburg, Germany
b University of Hamburg, Institute of Physical Chemistry, Grindelallee 117, 20146 Hamburg, Germany
c ams-OSRAM International GmbH, ams OSRAM Group, Leibnizstr. 4, 93055 Regensburg, Germany
d Helmholtz-Zentrum Hereon, Institute of Membrane Research, Max-Planck-Straße 1, 21502 Geesthacht, Germany
e Hamburg University of Technology, Institute of Advanced Ceramics, Integrated Materials Systems Group, Denickestraße 15, 21073 Hamburg, Germany

E-mail: g.schneider@tuhh.de

Abstract:

Additive manufacturing of shape memory polymers has gained tremendous interest in recent years due to their versatile potential applications in various industries, such as biomedicine and aerospace. However, the polymers' mechanical properties, specifically stiffness and strength, often hinder their use in mechanically demanding applications, e.g. as structural materials. In this work we produced nacre-inspired composites with a covalent adaptable network, which enables fast and mechanically lossless self-healing, reshaping, and shape memory capabilities. We demonstrate a novel direct write 4D-printing strategy to print smart, nacre-inspired, organic-ceramic composites based on alumina platelets and vitrimers with up to 3.3 and 26.7 times higher tensile strength and stiffness, respectively, in comparison to the pristine vitrimer. To the best of our knowledge, we introduce for the first time a single step 4D-printing process for nacre-inspired composites, that exploits suspension spreading to align micron sized alumina platelets along a common plane, and which utilizes solvent evaporation to induce polycondensation of the monomers resulting in a vinylogous urethane vitrimer after extrusion of the suspension. This work presents a facile, direct write 4D-printing strategy at ambient printing conditions, establishing a foundation for adaptive additive manufacturing of smart organic-ceramic composites of interest to various industries.

Keywords: additive manufacturing, colloids, vitrimers, evaporation induced polymerization, self-healing, shape memory materials, nacre-inspired composites

# 1. Introduction

Additive manufacturing of shape memory materials (SMM) has recently drawn extensive attraction with consequent expansion of its research field due to the versatile applicability of the process and its products, for instance in the fields of biomedicine[1,2] and aerospace engineering[3–5]. Since this processing strategy was introduced in 2014[6], different printing approaches have been studied[7]. To refer to all processes comprising printing of materials that exhibit additional shape transformation over time - which include printing of SMMs, the term "4D-printing" was established[7].

The fabrication of SMMs has already been applied to different material classes: metals[8], ceramics[9], composites[5] and polymers[10]. In fact, the latter is the most extensively studied material in the field of 4D printing research[7]. Many shape memory polymers (SMPs) can be processed with common additive manufacturing techniques, such as vat polymerization processes (stereolithography, digital light processing) or extrusion-based printing techniques (fused deposition modeling, direct writing)[7]. Moreover, the materials' properties - including mechanical properties, glass transition temperature ($T_g$), or actuation mode- can be tailored for the final application due to the great adaptability of polymeric materials. Furthermore, SMPs are generally low cost with further advantage such as low density, large deformability and fast response rates compared to other shape memory materials, like metallic shape memory alloys[10].

However, the permanent shape of conventional thermosetting SMPs, such as polyurethanes[3], is defined during printing. After processing, the permanent shape can no longer be adjusted to altering application conditions, thus reducing the overall versatility of the printed component. Additionally, such crosslinked polymers are not intrinsically recyclable, thereby contributing to economic and ecologic burden[10]. In this regard, a novel class of polymers, named vitrimers [11], appear as a sustainable opportunity, since they can be reshaped and recycled, even though they possess a permanently crosslinked network that makes them chemically resistant[12,13]. Due to an associative exchange mechanism, covalent bonds can be exchanged in the vitrimer upon heating, while keeping the number of crosslinking points constant in the polymer network[12]. These characteristic properties extend their applicability, presenting a versatile, economically and ecologically friendly polymer system. Vitrimers based on vinylogous urethane bonds are especially interesting for 4D-printing approaches, since their facile synthesis via condensation polymerization can proceed at ambient temperatures and without the presence of catalysts, while the polycondensation can be easily controlled by the concentration of the monomers in the printing suspension[12–14]. Moreover, the properties of vitrimers can be easily tailored by changing at least one of the monomers[12,14].

However, the mechanical properties of vitrimers (and polymers in general) -in particular hardness, stiffness, and strength- are often not sufficient to use them as structural materials. This drawback can be overcome by the addition of stiff and strong filler materials, leading to composites with improved mechanical properties[15]. Specifically, brick-and-mortar structures inspired by the natural hierarchical material, nacre, promise to yield hard, stiff, strong and tough composites[15]. Many nacre-inspired composites have already been presented in literature, which demonstrated promising mechanical performances[15–32]. However, common fabrication techniques for nacre-inspired composites, such as particle settling[29,33], layer-by-layer assembly[16,17,27,28,31] or freeze casting[18,22,26] limit the design flexibility to simple bulk components or thin planar films.

Here is where additive manufacturing has recently gained growing attention, since it allows producing complex shaped, nacre-inspired components[34]. To align the particles in a brick-and-mortar structure, mainly field-assisted particle orientation[35] and shear alignment induced within the printing nozzle during extrusion[36] have been used[34]. While field-assisted additive manufacturing techniques require field-responsive particles and custom printing equipment, shear within the nozzle orients particles inevitably in accordance with the nozzle's cross section[37,38], thus causing partial misalignment with regard to a targeted planar brick-and-mortar structure. However, spreading low viscous suspensions on substrates was reported to produce well-oriented, nacre-inspired composite thin films[32,39], thus comprising an alternative shear alignment approach that avoids the aforementioned particle misalignment induced by the nozzle.

In this work, we successfully applied this alignment strategy for the direct writing of low viscous suspensions to produce nacre-inspired, smart, ceramic-organic composites. In the following, we introduce a facile 4D-printing approach based on direct writing to print nacre-inspired composites consisting of a vinylogous urethane vitrimer matrix and functionalized, micron-sized alumina platelets. In this 4D-printing process, particles align after extrusion while solvent evaporation additionally induces polycondensation to form the vitrimer matrix, which binds covalently to the aligned and functionalized platelets. Furthermore, we show how the composite's nacre-inspired structure increases its strength and stiffness up to 3.3 and 26.7 times, respectively, in comparison to the pristine vitrimer matrix without reinforcement. Moreover, the mechanical behavior of the printed composite can be tuned by adjusting the curing time, so that a response ranging from ductile or elastoplastic to brittle can be induced. Moreover, we finally demonstrate that the composite exhibits self-healing, reshaping and shape-memory capabilities, thus enabling the fabrication of smart, complex geometries with a planar brick-and-mortar structure, which can be reprogrammed for multiple uses. To the best of our knowledge, this is the first report of a 4D-printing approach based on particle shear alignment during directed self-assembly while printing (additive manufacturing combined with self-assembly) and simultaneous evaporation induced polymerization. This establishes a facile direct write printing strategy at ambient printing conditions, which paves the way to the adaptable production of smart composites for various industries, such as biomedicine and aerospace.

## 2. Materials and Methods
### 2.1. Particle functionalization

*α*-alumina platelets with $d_{50} = 2\ \mu m$ and thickness of ~0.04 µm (determined by supplier), resulting in an aspect ratio of ~50, were purchased from Kinsei Matec Co.,LTD (Japan). These particles were functionalized with (3-aminopropyl)trimethoxysilane (APTMS, Merck KGaA, Darmstadt) based on a procedure previously reported in [40,41]. In summary, 1 g of *α*-alumina platelets were dispersed in 18 mL ethanol and ultrasonicated for 15 min. Afterward, the suspension was heated to 60°C. An amount of 0.185 g APTMS as well as 2 mL of water were added while magnetically stirring the suspension. The amount of APTMS was chosen based on an estimate for the amount required to form a theoretical multilayer structure with up to ten APTMS layers on the particle's surfaces (10eq APTMS). The resulting suspension was kept at 60°C with constant stirring for 30 min. Subsequently, the particles were washed three times with ethanol and finally dispersed in ethanol with a concentration of 40 mg mL$^{-1}$.

### 2.2. Suspension preparation

The suspensions used for printing were always freshly prepared before printing. The prepared batches were used up to 30 min for printing and switched at least every 30 min to avoid needle clogging as well as alterations of the printing conditions (see Results and Discussion). Each batch of about 1 mL suspension's volume in total contained 40 mg mL$^{-1}$ APTMS-functionalized alumina platelets. Higher particle concentrations caused needle clogging, thus they were avoided.

To form a vitrimer-matrix between the platelets upon solvent evaporation, 2,2'-(ethylenedioxy)diethylamine (Sigma Aldrich) and propane-1,2,3-triylbis(3-oxobutanoate) (trifunctional acetoacetate) were added to the suspension. The trifunctional acetoacetate was synthesized via acetoacetylation of glycerol. The acetoacetylation was carried out according to the synthesis route of Wu et al.[42] by mixing glycerol (10.00 g, 0.108 mol (99.5%, ROTH) and *tert*-butylacetoacetate (54.10 g, 0.342 mol, (98%, Alfa Aesar)) in a 250 mL flask. The mixture was heated at 130 °C for 3 h, while the byproduct *tert*-butanol was removed by distillation during the reaction. The temperature in the distill head was usually between 75–90 °C. When the temperature dropped to 35 °C, the remaining acetone, 2,2,6-trimethyl-1,3-dioxin-4-one, and dehydroacetic acid were removed in vacuum at 135 °C and 0.1 mbar for 30 minutes. The resulting product is a slightly yellowish liquid containing the desired monomers.

The total monomer content in the suspensions was varied (10, 20, 30, 40, 50, 60 mg mL$^{-1}$) to assess the monomer printability whilst assuring the fitting between polymerization and printing times. The maximum printable concentration was 60 mg mL$^{-1}$, because further increase in total monomer concentration clogged the needle tip frequently. The individual monomer content in the suspensions was defined by the total monomer content, but also by the ratio of the functional groups, i.e., the acetoacetate to amine ratio $R$, which was kept constant at 0.8 for all suspensions. With $R = 0.8$ an amine-excess was present, which formed a vitrimer network, exhibiting fast transamination reactions and enabling fast self-healing capabilities[14].

### 2.3. Direct Writing & Post processing

For printing, the extrusion-based BioX$^{TM}$ printer from CELLINK was used. It was equipped with a pneumatic print head that fitted 3 mL cartridges. We filled the cartridges with 1 mL of the suspensions that were prepared as described in the previous section. At least every 10 min the cartridges were agitated between printing steps to maintain homogeneity and prevent particle settling in the cartridge. Needles from Nordson Corporation with 27 gauge and 9 mm length were used. Due to the small yield stress ($\tau_0 = 18$ mPa), low viscosity (the maximum apparent viscosity, η, measured at $\dot{\gamma}=0.01$ s$^{-1}$ shear was η = 140 mPa s) and shear thinning behavior (flow behavior index n = 0.29) of the suspension, the minimum applicable pressure of 1 kPa was sufficient for a continuous extrusion (**Figure S1**). A writing speed of 30 mm s$^{-1}$ was applied, because it was found to produce the thinnest continuous lines without disruptions. Various substrates with different contact angles between suspension and substrate were tested for printing (glass, silanized glass, poly(vinyl alcrohol) (PVA), poly(methyl methacrylate) (PMMA), and Parafilm). Parafilm was identified to be the best suitable (details in the section 3.1.2.) and used for successive experiments in this study. The printed, dried samples were released from Parafilm via quenching in liquid nitrogen. The sub-millimeter thin samples were later post-cured on a heating plate at 80 °C at different dwell times (3, 12, 18 and 24 h). For tensile testing 50 mm long bars were printed, consisting of 30 printed layers.

### 2.4. Characterization

The surface functionalization of the particles was characterized via Fourier-transform infrared spectroscopy (FTIR, Tensor II, Bruker). KBr pellets were prepared using 0.3 wt% of dried particles, which were measured in transmission in a frequency range of 4000 to 400 cm$^{-1}$. The zeta-potential was assessed with a Zetasizer Nano ZS at 25 °C (Malvern Panalytical Ltd.), which gave further information about the platelets' surface properties as well as the suspension's stability. The stability of the colloidal suspensions was additionally examined via sedimentation experiments similar to a study previously reported[43]. Polymer cuvettes with cuboid shape were filled with 3 mL of the suspensions, which were manually agitated and sonicated for 5 min beforehand to ensure an initial homogenous particle distribution. A backlight was placed behind the cuvettes and a camera (DMC-FZ300, Panasonic) was mounted in front, which captured the cross section of the cuvettes. The camera took an image every minute for a total duration of 90 min. The flow behavior of the suspensions was investigated with via rotational rheometry (Kinexus Pro, Netzsch) at 25 °C, with a double gap bob suitable for low viscous fluids and with a gap size of 1 mm.

The printed samples were analyzed with a scanning electron microscope (SEM, Zeiss Supra VP55, Carl Zeiss AG) applying low accelerating voltage of 1.5 kV at a working distance of about 5 mm. The particle orientation in the samples was assessed based on SEM cross sections and by applying the "Directionality" plugin in ImageJ with a 5x5 Sobel-operator. We used attenuated total reflection Fourier-transform infrared spectroscopy (ATR-FTIR, Bruker FT-IR Vertex 70) to characterize the cross-linked vitrimer in the printed composite and thermogravimetric analysis (TGA, STAR$^e$ System, METTLER TOLEDO) with a 10 K min$^{-1}$ heating ramp up to 900 °C in synthetic air to determine the weight ratio between the organic and the inorganic phase.

As described in the previous section, composite samples were printed for tensile testing. The printed samples had a width of about 2 mm and a height of about 0.2 mm. No post-processing was performed after printing (apart from a curing step) to avoid introduction of defects. Thus, the samples' cross sections were of a circular segment shape and their surface contained undulations. Hence, to be able to assess the cross-section area accurately, the samples' topographies were acquired with an optical surface metrology instrument (Alicona G4, Infinite Focus). Reference vitrimer samples with equivalent dimensions were casted in molds with 2 mm width. Tensile tests were performed (zwickiLine, Zwick Roell) with the printed and casted bars (bars were about 25 mm long, details about printing process for the composite bars can be found in the previous section). The reported mechanical properties are the averaged values of 4 measured samples (except for values of samples printed with other than 60 mg mL$^{-1}$ total monomer concentration, more details in the Results and Discussion section). The mechanical, and self-healing capabilities of the samples were evaluated by dynamic mechanical analysis (DMA, MCR 502 (Anton Paar GmbH, Graz, Austria). An 8 mm plate-plate geometry and a heat chamber with flooded nitrogen atmosphere was used for oscillatory shear experiments. The gap between the lower and the upper plate was usually set to 0.5 mm and the normal force was 1 N. Temperature-dependent measurements were performed at a constant shear strain of 0.1 % in the temperature range of 0–170 °C. Stress relaxation experiments were carried out with a shear strain of 0.5 %, and the relaxation modulus was recorded as a function of time at 130 °C. The glass transition temperature, $T_g$ , of the vitrimer and composites was assessed via differential scanning calorimetry (DSC, DSC 204 F1 Phoenix). Four heating cycles under nitrogen were performed on each sample with a rate of 10 K min$^{-1}$: from -50 °C up to 80°C the first three scans and a fourth scan up to 150 °C. The reshaping capability of the printed composites was tested on a heating plate at 80 °C. Shape-memory experiments were performed with the same samples on a heating plate at 60 °C and by exposing them to body temperature (~37 °C).

## 3. Results and Discussion
### 3.1. One-pot printing approach

Our printing process is based on direct writing of low viscous suspensions (**Figure 1A**). The suspension contained micron sized α-alumina platelets ($c_{alumina}$ = 40 mg mL$^{-1}$). Additionally, trifunctional acetoacetates and diamines were dissolved in ethanol to form a vitrimer network between the platelets (**Figure 1B**). The bare surface of the alumina platelets only allows for weak electrostatic interactions with the vitrimer matrix. Hence, the platelets' surface was modified by introducing a silane-derived amine ligand, (3-aminopropyl)trimethoxysilane (APTMS). The silane anchoring group covalently binds to the alumina surface, while the amine group acts as chemical anchor point for the vitrimer matrix (Figure 1B). The initial concentration of the monomers in the suspension was kept at a maximum concentration of 60 mg mL$^{-1}$ to significantly reduce polycondensation reactions, such that the suspension could be stored in cartridges for printing for at least 30 min (more details in the following section).

Upon deposition on the substrate, the suspension immediately spread. In our printing process, spreading of the suspension shall induce a shear flow, which aligns platelet particles concomitantly with the particles' self-assembly after extrusion along a common plane (Figure 1C). The low monomer concentration in the printing suspension reduces the polymerization kinetics to virtually zero (more information is given later). Upon extrusion, however, the solvent evaporates, which increases the monomer concentration, thus inducing polycondensation (Figure 1D). Consequently, monomers in the suspension form a crosslinked vitrimer network and acetoacetates additionally form covalent bonds with the amine moieties on the particles' surface, which assures a strong binding at the interface. Once the solvent is fully evaporated, a nacre-inspired composite is formed (Figure 1E). In the following subsections, we will present and discuss the aforementioned characteristics of the printing process in more detail, i.e. the composition and stability of the suspension, platelets' shear alignment, formation of the vitrimer network as well as the produced nacre-inspired structure.

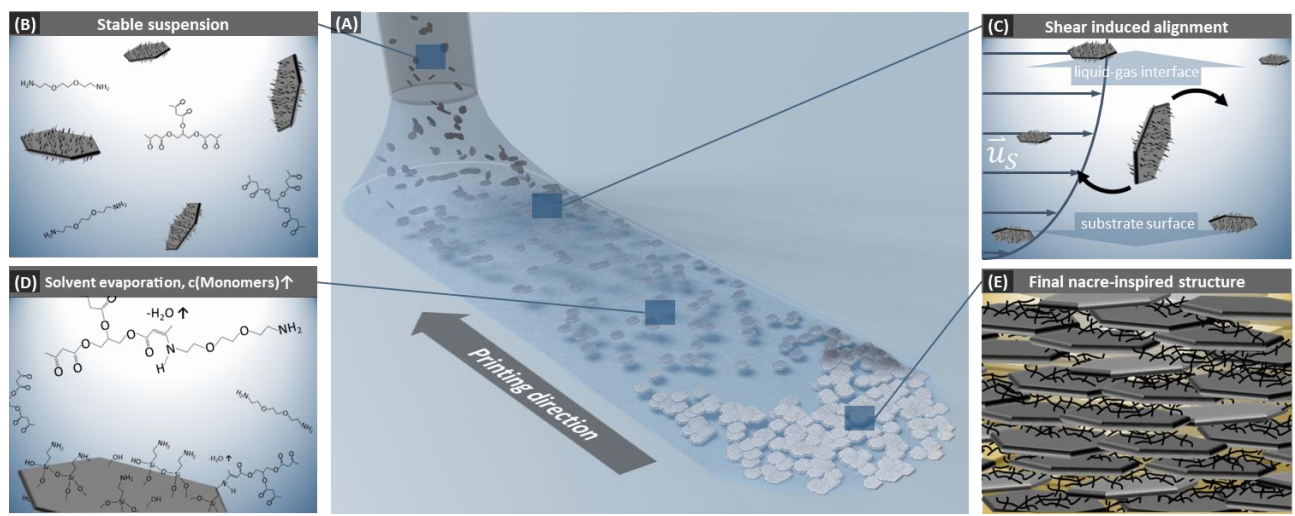

Figure 1: Schemes of (A) the 4D-printing approach. (B) The suspension in the cartridge consists of APTMS-functionalized alumina platelets, trifunctional acetoacetates and diamines with ethanol as solvent. (C) Platelets align after extrusion due to shear flow induced by suspension spreading. (D) Solvent evaporation induces vitrimer formation by polycondensation and covalent bonds between organic matrix and inorganic filler (E) After complete solvent evaporation, a nacre-inspired composite is formed.

### 3.1.1. Particle functionalization and suspension characteristics

Prior to printing, the alumina platelets were functionalized with 10eqAPTMS, which refers to an estimated amount required to theoretically form ten layers of APTMS on the platelet's surface (further details in the Methods section). FTIR measurements in transmission showed significant differences in the spectrum of the $Al_2O_3$@10eqAPTMS sample compared to a bare $Al_2O_3$ reference sample, which comprised the same platelets, but non-functionalized (**Figure 2A**). Peak broadenings at 470 cm$^{-1}$ and 580 cm$^{-1}$ as well as a characteristic absorption at 1230 cm$^{-1}$ was ascribed to Si-O-Si bonds, also present in the APTMS reference spectrum (Figure 2A)[44–46]. Furthermore, the absorption peak at 1385 cm$^{-1}$ in the $Al_2O_3$@10eqAPTMS and APTMS reference spectra corresponds well with C-H bending modes in -$CH_3$ groups[47]. Therefore, the functionalization with 10eqAPTMS was deemed successful and used throughout this work.

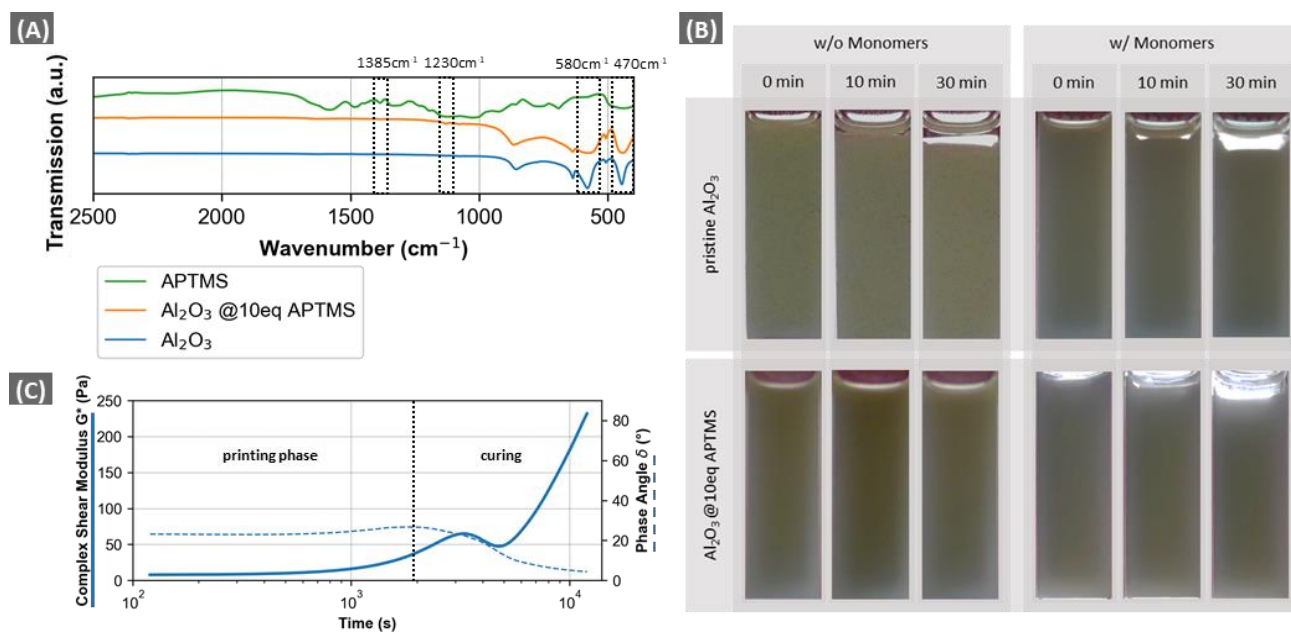

**Figure 2:** (A) FTIR transmission spectra of non-functionalized particles ($Al_2O_3$), functionalized particles ($Al_2O_3$@10eqAPTMS) and APTMS. (B) Photographs of the sedimentation experiments taken at the beginning, after 10 min and 30 min settling time for pristine and functionalized particles, with and without monomers present in the suspension (w/ Monomers and w/o Monomers, respectively). (C) Vulcanization curve for suspension initially containing 60 mg mL$^{-1}$ monomers. The dashed vertical line separates the duration used for printing from the curing phase.

To define suspension storage requirements in cartridges during printing, sedimentation experiments were performed by investigating the particle settling in the low viscous suspensions under different conditions. Without monomers being present in the suspension, the functionalized particles ($Al_2O_3$@10eqAPTMS) remained completely stable for at least 90 min, i.e. no clarification front could be seen on top of the cuvette. In the cuvette containing pristine $Al_2O_3$ (non-functionalized particles), however, a clarification initiated at the meniscus after 10 min. The greater stability of the suspension containing $Al_2O_3$@10eqAPTMS platelets could additionally be validated with zeta potential measurements (**Table 1**). While pristine $Al_2O_3$ platelets showed a zeta potential of 8.9 ±0.3 mV in ethanol without monomers, thus exhibiting little electrostatic repulsion with propensity for aggregation and settling, the $Al_2O_3$@10eqAPTMS platelets presented a zeta potential of 60.9 ± 1.6 mV. Here, protonation of the amine moieties on the platelet's surface caused the comparatively large, positive zeta potential. When monomers were added to the suspensions, both suspensions showed an initial clarification after 10 min (Figure 2B). After 30 min, a clarification front with a thickness of about 2 mm was formed in both samples, which corresponded to 7 % of the total filling height (28 mm, Figure 2B). This observed reduction in suspension stability for the functionalized platelets agreed with zeta potential measurements (Table 1). Both suspensions presented a negative zeta potential, with absolute values smaller than 40 mV, which is interpreted as insufficient repulsion to prevent particle agglomeration and settling within 30 min. The addition of the monomers, especially of diamines, increased the pH to >10 (Table 1). The basic environment being responsible of the deprotonation of the previously protonated -$NH_3^+$ moieties on the platelets' surface. Additionally, hydroxyl groups being present on both, pristine $Al_2O_3$ and $Al_2O_3$@10eqAPTMS platelets' surfaces, potentially formed the negative zeta potential[48]. Based on these findings the printing suspensions were homogenized in 10 min intervals to ensure a homogenous particles distribution within the cartridge during printing.

**Table 1: Zeta-potential measurements of functionalized and non-functionalized particles in EtOH with monomers and without monomers present (w/Monomers, w/o Monomers)**

|  |  | Zeta-potential (mV) | pH |
|---|---|---|---|
| Pristine $Al_2O_3$ | w/o Monomers | 8.9 ± 0.3 | 7.1 |
|  | w/ Monomers | -38.8 ± 1.6 | 10.5 |
| $Al_2O_3$@10eqAPTMS | w/o Monomers | 60.9 ± 1.6 | 7.9 |
|  | w/ Monomers | -20.3 ± 1.2 | 10.3 |

Next to platelet settling, the tailoring of the onset of curing of the monomers in the suspension was crucial for storing the suspension during printing. To allow for storing in a cartridge during printing, the reaction rate of the polycondensation was significantly reduced by diluting the monomers' concentration in the suspension. The maximum total monomer concentration being used for printing was 60 mg mL$^{-1}$, which led to an onset of curing after 30 min, as can be seen by the decrease of the phase angle in the oscillatory measurement (Figure 2C). After 30 min, both, phase angle and complex shear modulus of the suspension progressed in a typical vulcanization curve[49]. For monomer concentrations <60 mg mL$^{-1}$, the curing onset was further delayed, since the probability of a polycondensation reaction becomes smaller with decreasing monomer concentrations. As a consequence, we prepared single 1 mL suspension batches on demand and stored them up to 30 min in the printing cartridges. At least every 30 min we switched the cartridge between printing steps to maintain consistent printing conditions.

### 3.1.2. Printing nacre-inspired organic-ceramic composites

Suitable substrates for the printing strategy were chosen by prinitng single lines with a suspension containing 40 mg mL$^{-1}$ of functionalized platelets and a total monomer concentration of 10 mg mL$^{-1}$ (**Figure 3A**). Lines printed on glass, silanized glass, PMMA, and PVA formed line cross sections with an inhomogeneous particle distribution: Platelets (white domain in lines, Figure 3A) stayed within a confined space, while the organic phase spread over the entire wetted area (transparent domain of the lines, Figure 3A). The substrates induced complete wetting in proximity to the original contact lines, which led to an extensive spreading of the solvent on the substrate, within which only the organic phase migrated, while the platelets remained in a confined area at the middle (**Figure S2**). On Parafilm, however, such particle confinement was not observed, and a homogenous line could be printed (Figure 3A). The wetting of the suspension on Parafilm was still high (contact angle of suspension on Parafilm $\theta_{Susp} = 16.7 \pm 0.3$ °; ethanol on Parafilm $\theta_{EtOH} = 22$ °[50]), but no total wetting was observed as for the other substrates, thereby the spreading of the solvent did not confine platelets. Instead, a homogeneous spreading of both platelets and organic phase is observed over the entire printed line cross-section (Figure S2). Since an homogenous materials distribution in the printed structure is required for printing a composite with a defined content of organic and inorganic phase, Parafilm was chosen as the substrate for subsequent printing experiments.

Line stacks of up to 30 layers were printed for mechanical analysis, which will be discussed in the following section (3.2.). SEM cross sections of the printed bars showed a composite's particle alignment along a common plane parallel to the substrate surface (Figure 3B), which is essential for the fabrication of nacre-inspired structures. By applying the directionality function in ImageJ on the SEM micrographs, the individual particle orientation could be determined with a Sobel-operator and each orientation angle was assigned to a certain colour. The overlay of the platelets' edges with the assigned colour led to an overall blue coloration of the SEM micrograph, indicating alignment along the horizontal plane within approximately ±14°, which was determined with the standard deviation of the Gaussian fitted to the particle orientation distribution (Figure 3B). We assigned this alignment to the initial spreading of the suspension after extrusion, which induces a shear flow that acts upon the particles, as observed in other studies of thin films[32,39]. To the best of our knowledge, a shear flow alignment outside of the needle (nozzle) has not yet been applied in additive manufacturing processes of multi-layered macroscale composites, such as the one demonstrated here. The shear rate in the spreading suspension can be estimated by $\dot{\gamma}_S = u_S/H$, where $u_S$ is the spreading velocity of the suspension on the substrate and $H$ the height of the deposited wet line[32]. In our experiments the maximum height of the printed lines was $H<1$ mm, and the spreading velocity $u_S$ is about 0.2 m/s for ethanol on Parafilm[50]. Hence, an estimate for $\dot{\gamma}_S$ is ~200 s$^{-1}$ considering our printing conditions. Strain rate sweeps demonstrated that our low-viscous suspension used for printing is highly shear thinning, with a measured flow index of $n = 0.29$, and that the shear viscosity reaches a plateau at shear rates >100 s$^{-1}$ (Figure S1). This indicates that the majority of platelets aligns at such or higher shear rates, and thus they also aligne in the spreading suspension after extrusion from the printing nozzle.

Moreover, the natural self-assembly ability of the platelets, when present in stable suspensions, can also contribute to their alignment[51]. However, under just evaporative self-assembly conditions, a fast solvent evaporation potentially leads to undesired inhomogeneities within the structure. In contrast, the fast printing process reported here allows to obtain well-aligned nacre-inspired composites within 30 min. Furthermore, the timescale for the platelet alignment can be estimated by $t_{shear} = 1/\dot{\gamma}_S$ [32]. Hence, considering the alignment induced by the spreading shear flow, in our printing process the majority of platelets were aligned immediately after extrusion ($t_{shear}$ = 5 ms), which is essential for the printing strategy to form a nacre-inspired platelet alignment prior polycondensation of the vitrimer network. Coming back to the common evaporative self-assembly example: under such conditions, the polycondensation leading to the vitrimer network would be finished prior to the particles' self-assembly by evaporation (which often takes days), hindering the platelets mobility and alignment.

ATR-FTIR spectra of the printed composites, the reference spectra of the vitrimer, and the $Al_2O_3$ platelets confirmed the formation of a vitrimer network during printing (Figure 3C and Figure S4). The characteristic C=O stretching vibration band at 1647 cm$^{-1}$ and C=C stretching vibration band at 1590 cm$^{-1}$ were identified for the vinylogous urethane bonds[12], and are present in the printed composite's spectrum as well as in the vitrimer's reference spectrum. The ATR-FTIR spectrum for alumina platelets showed a broad absorption band from 860 cm$^{-1}$ towards lower wavenumbers, which is assigned to Al-O bonds[52]. The broad band absorption could also be found in the composite as an overall increase in absorption in this wavenumber range, confirming the presence of both phases in the composite. Furthermore, temperature-dependent DMA confirmed the formation of cross-linked materials (Figure S5).

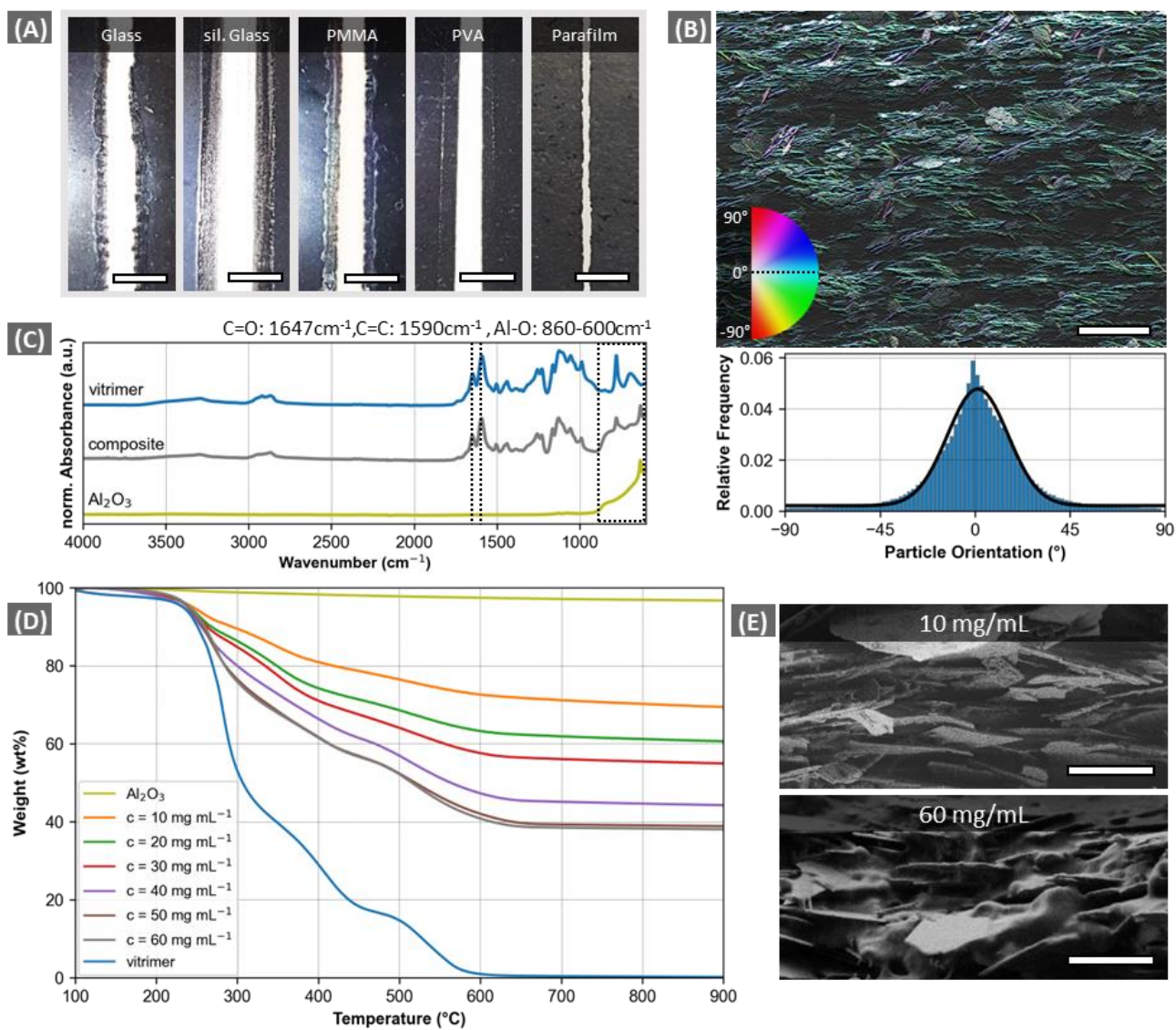

**Figure 3: (A)** Photographs of single line prints on various substrates (From left to right: glass, silanized glass, poly(methyl methacrylate), polyvinyl alcohol, Parafilm) **(B)** Top: SEM micrograph of composite cross section (30 layers printed with 10 mg mL$^{-1}$ monomer concentration). The coloration of the particle edges marks the relative particle orientation angle towards the horizontal plane (See color wheel for specific orientation angle). Scale bar: 20 µm. Bottom: Histogram and corresponding gaussian fit for the determined particle orientation angles in the SEM cross section (Particle orientation of 0° corresponds to horizontal alignment). **(C)** ATR-FTIR spectra of the vitrimer, composite (60 mg mL$^{-1}$, 24 h at 80° C) and alumina platelets. **(D)** TGA measurements in synthetic air for alumina, composites printed with various monomer concentrations (10, 20, 30, 40, 50, 60 mg mL$^{-1}$) and the vitrimer. **(E)** SEM cross sections of samples printed with 10 mg mL$^{-1}$ and 60 mg mL$^{-1}$ monomer concentrations (top and bottom).

The nacre-inspired structure was highly porous when being printed with the lowest monomer concentration $c = 10$ mg mL$^{-1}$ (Figure 3E), because the volume of the formed vitrimer was too little to fill the gaps between platelets. Hence, the monomer concentration was increased in 10 mg mL$^{-1}$ steps up to a maximum concentration $c = 60$ mg mL$^{-1}$, aiming for a final dense composite (Figure 3E).

TGA analyses of the printed composites confirmed the tuneability of the composition by adjusting the initial monomer concentration in the suspension (Figure 3D, **Table 2**). A maximum deviation of 7 wt% for the produced ceramic content in the composite from the targeted ceramic content, was measured for the 10 mg mL$^{-1}$ monomer concentration (Table 2). In composites with such low vitrimer content, loose particles without sufficient contact to the vitrimer network may have been removed during sample handling, thus causing a lower ceramic content in the printed structures than anticipated. For monomer concentrations higher than 10 mg mL$^{-1}$, the final vitrimer content in the composite increased accordingly (max. deviation 3 wt%), thus proving the process stability of the printing strategy (Table 2).

**Table 2: Targeted ceramic content and measured ceramic content (via TGA) in samples printed with various monomer concentrations in the suspension (10, 20, 30, 40, 50, 60 mg mL$^{-1}$) as well as the deviation of the ceramic content in the printed samples from the targeted ceramic content.**

| Suspension's monomer concentration (mg mL$^{-1}$) | 10 | 20 | 30 | 40 | 50 | 60 |
|---|---|---|---|---|---|---|
| targeted ceramic content (wt%) | 80 | 67 | 57 | 50 | 44 | 40 |
| Measured & drift corrected ceramic content (wt%) | 73 | 64 | 58 | 47 | 42 | 41 |
| Deviation from the target | 7 | 3 | -1 | 3 | 2 | -1 |

### 3.2. Mechanical characterization

Printed bars of the nacre-inspired composite consisting of 30 printing layers and with a length of about 25 mm were mechanically tested under tensile mode (**Figure 4A**). The profile of the cross section of the bars was circular segment-shaped (Figure 4B). The as-printed composites mainly exhibited similar or lower tensile strengths than the reference vitrimer sample (Figure 4C). Only the sample printed with a monomer concentration of $c = 50$ mg mL$^{-1}$ reached higher tensile strength than the reference (8.7 MPa vs. 3.0 MPa for the vitrimer), but the maximum strain was significantly reduced for the printed composite (29 %. vs. 101 % for the vitrimer). In the as-printed composites the platelets did not reinforce the vitrimer, but formed defects that caused failure at lower strains.

To take advantage of the nacre-inspired structureand activate the functionalized-interfaces-strengthening mechanism, samples were heat treated at 80 °C for 24 h. After the additional curing step, the composites became stiffer than the cured vitrimer references and composites printed with monomer concentrations >30 mg mL$^{-1}$ also possessed a higher tensile strength than the references (>16.1 MPa tensile strength for 24 h-cured composites, Figure 4D). The additional crosslinking of the vitrimer network reduced the strain at failure to 10.0 % for the vitrimer reference sample and to 0.7 % on average for the 24 h-cured composites. The tensile strength of the composites increased gradually with increasing monomer concentration, from 7.8 MPa for the 10 mg mL$^{-1}$ composite up to 73.0 MPa for the cured composite printed with the highest printable monomer concentration (60 mg mL$^{-1}$, Figure 4D). With increasing monomer concentration, the pores between platelets were still filled further, thus expanding the interface between the organic and inorganic phases and thus reducing the number of defects, i.e. pores, which improved the transfer and distribution of load from the soft vitrimer to the stiff ceramic platelets. Even though a high ceramic content is favored in nacre-inspired structures, the reduction of open pores within the composites was a key to improve the mechanical behavior, i.e. stiffness, strength, and toughness[16,17]. Only the 50 mg mL$^{-1}$ sample presented a lower strength than a sample printed with less monomer concentration, i.e. 40 mg mL$^{-1}$. Here, fewer cross links in the vitrimer network may have formed during curing, which lead to reduced strength, but to a greater strain at failure (1.2 % for $c$ = 50 mg mL$^{-1}$, Figure 4D).

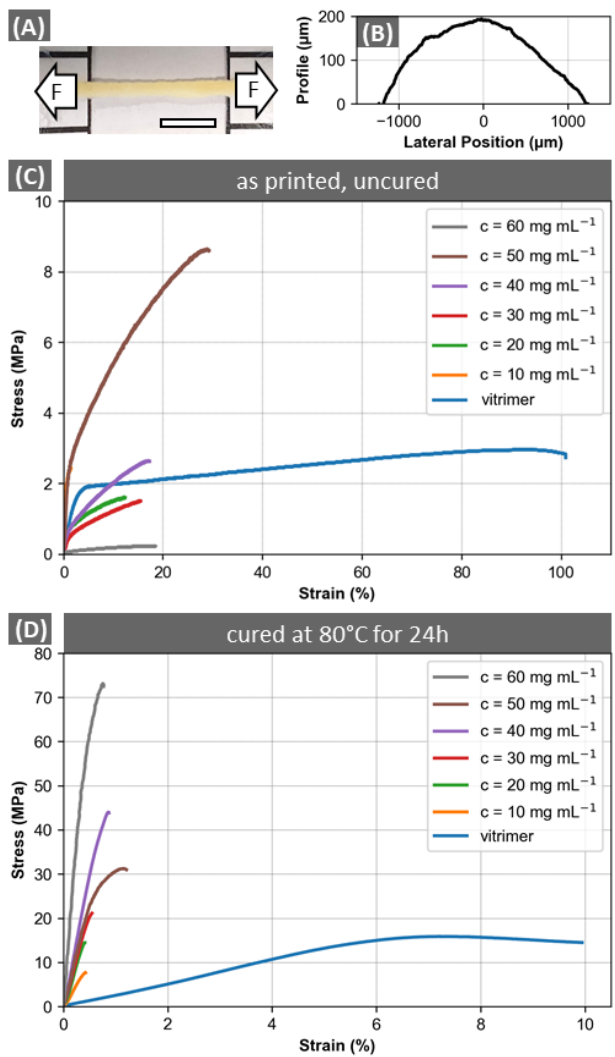

**Figure 4: (A) Photograph of printed sample and prepared for mechanical testing. Scale bar: 5 mm. (B) Exemplary profile of sample's cross section, which was tested in tensile mode. (C) Stress-strain curves of samples printed with**

**various monomer concentrations (10, 20, 30, 40, 50, 60 mg mL$^{-1}$), which were tested under tension. (D) Stress-strain curves of samples printed as in (A) but additionally being cured at 80 °C for 24 h prior tensile testing.**

In nacre-inspired composites loads must be transferred from the organic to the inorganic phase to strengthen the material[16], hence next to the mechanical performance of the organic matrix and inorganic reinforcement, understanding the mechanical behavior of the interface is essential. To evoke and investigate different mechanical behavior in the composites as well as the mechanical performance of its three components, we cured the printed composites for various durations (0 h, 3 h, 12 h, 18 h and 24 h). By reducing the curing duration from 24 h to 18 h and 12 h the composite turned from a rather brittle and stiff material into an elastoplastic composite (Figure 5A). After 3 h curing it remained ductile as in the uncured state, but presented a higher strength and lower strain at failure (7.2 ±2.7 MPa, 11.9 ± 2.5%) than without curing (5.1 ±5.0 MPa, 51.5 ± 23.4%, Figure 5A, B). In general, increasing the curing duration reduced the strain at failure, but increased the tensile strength and stiffness of the composite (Figure 5B). This is associated to a stiffer and stronger vitrimer matrix, directly bonded to the functionalized platelets. While SEM micrographs confirmed a ductile failure of the vitrimer matrix in uncured samples, brittle failure could be seen in the 24 h-cured composites, as indicated by a rather planar fracture surface (Figure 5C). Additionally, exposed platelet surfaces, i.e. platelet surfaces not being covered by the vitrimer, could be found in SEM micrographs of the fracture surface for the 24 h-cured composites (Figure 5C). This indicates an interfacial failure of the vitrimer-functionalized platelet when loaded under tension, because the vitrimer network became stronger than the interface itself. Fracture surfaces of 12 h-cured samples confirmed the elastoplastic material behavior observed in the stress-strain curves (Figure 5A), achieved by nacre-like toughening mechanisms (Figure 5C): top-view SEM micrographs portray the presence of crack deflection - an extrinsic toughening mechanism-, while the SEM cross section of the fracture surface presented platelets still being covered with the vitrimer after fracture, thus indicating ductile platelet pull-out as an additional intrinsic toughening mechanism[15]. Hence, the 12 h-cured samples allowed the occarrance of extrinsic and intrinsic toughening mechanisms, which are essential to form damage tolerant composite materials[15].

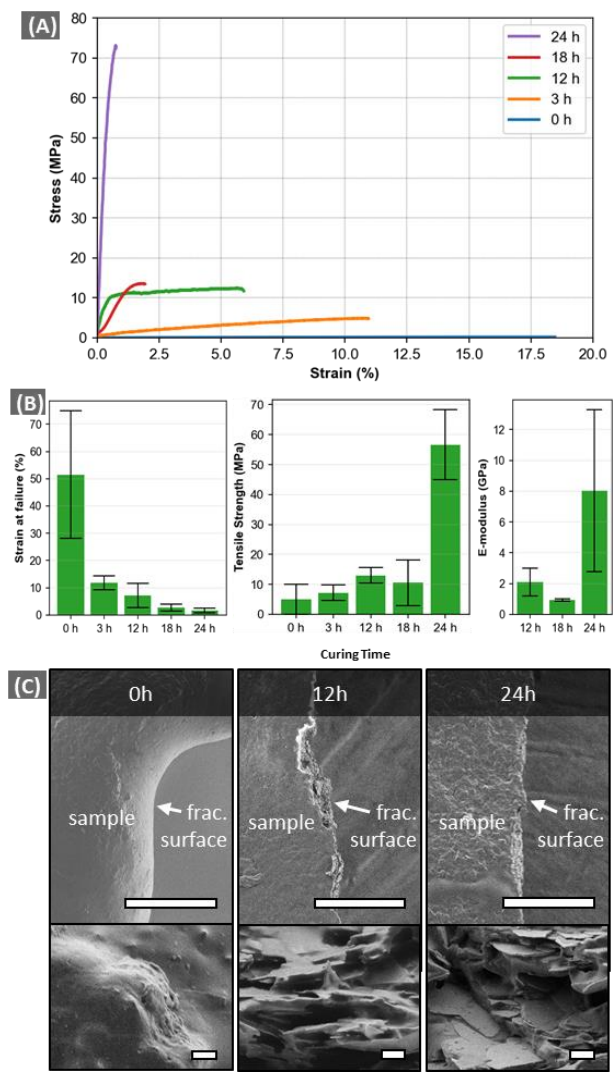

**Figure 5:** (A) Stress-strain curves of exemplary samples tested under tension after printing them with a monomer concentration of 60 mg mL$^{-1}$ and curing them for different durations (0, 3, 12, 18, 24 h at 80 °C). (B) From left to right: Average strain at failure, tensile strength and Young's modulus determined for samples prepared as in (A). Error bars indicate the standard error of four measurements. (C) SEM micrographs of fracture surfaces in top view (top row) of samples that were tensile tested along horizontal axis and SEM micrographs in cross-sectional view (bottom row) for the same tensile tested samples. The samples were heat treated for different durations (0, 12 and 24 h at 80 °C). Scale bars: 100 μm and 1 μm for top and bottom row, respectively.

In comparison to the 24 h-cured vitrimer reference sample, the 24 h-cured composite possessed a 3.3- and 26.7-fold strength and stiffness, respectively (**Table 3**). This increase is ascribed to the nacre-like reinforcement mechanism based on load transfer from the soft organic to the stiff inorganic phase, when loaded under tension. Other reinforcement mechanisms, such as a higher degree of polymer curing in the presence of particles, as it has previously been reported[28], is excluded, since the $T_g$ for both materials remained at 30 °C after curing (Figure S4). As mentioned before, the maximum strain was reduced by the incorporation of platelets: from 9.0 ± 8.1% for the vitrimer to 1.7 ± 0.9%. Here, the high standard deviation for the vitrimer is relatd to the presence of pores formed during sample preparation. While composite samples were printed, the vitrimer samples were casted. Hence, water formed during polycondensation could escape the single µm-thin printing layers in the composite, but pores filled with water vapor in the casted vitrimer samples could not escape the samples, thus creating defects. This demonstrates another advantage of our printing process, the Additive Manufacturing combined with Colloidal Self-assembly (AMCA), for the production of nacre-inspired composites. After 12 h curing, the strain at failure and tensile strength of the composite was slightly reduced in comparison to the 24h-cured vitrimer, while the stiffness of the composite was seven times higher (Table 3). This illustrates the great potential to improve and tune the mechanical performance by introducing a nacre-inspired structure, i.e. increasing stiffness but also toughness, which are usually mutually exclusive mechanical characteristics[15].

**Table 3: Mechanical properties (strain at failure, tensile strength, Young's modulus) of the vitrimer as well as nacre-inspired composites, which were printed with a total monomer concentration of 60 mg mL$^{-1}$ and subsequently cured for 12 and 24 h at 80 °C.**

|  | Strain at failure (%) | Tensile Strength (MPa) | E-modulus (GPa) |
|---|---|---|---|
| Vitrimer (24 h cured) | 9.0 ± 8.1 | 17.0 ± 5.2 | 0.3 ± 0.1 |
| Composite (24 h cured) | 1.7 ± 0.9 | 56.7 ± 11.7 | 8.0 ± 5.3 |
| Composite (12 h cured) | 7.1 ± 4.4 | 13.0 ± 2.7 | 2.1 ± 0.9 |

**3.3. Self-healing, reshaping and shape memory capabilities**

Previous studies have already presented the capabilities of vitrimers to reshape, to memorize shape, and to self-heal [12,13]. To prove and investigate the latter for our printed composites, relaxation experiments were performed with a 24 h-cured composite sample as well as a 24 h-cured vitrimer and an epoxy sample as a reference at 130 °C (**Figure 6A**). As can be seen in the relaxation curves for the vitrimer and the composite, the materials relaxed after an initially applied strain, whereas the normalized relaxation modulus for the epoxy reference remained at about 100 %. A total relaxation was achieved within ~3 min in the vitrimer as well as in the composite sample, thus illustrating that both materials possess self-healing capabilities. In this time period, associative exchange reactions based on transamination reactions reduce the loading by forming an adapted vitrimer network[12]. To quantify the self-healing capabilities, the time between initial straining and relaxation to 37 % (1/e) of the normalized relaxation modulus was determined according to Maxwell's model of viscoelastic fluids[13]: In our experiment, the vitrimer and composite relaxed within 13 and 15 s, thus exhibiting fast self-healing capabilities[12]. As shown by DSC measurements (Figure S4), the *Tg* of the 24h-cured vitrimer and the composite did not change when exposed to the healing temperature (130 °C) *e.g.*

during self-healing, hence the vitrimer network was not significantly affected by this additional heat exposure. To test the mechanical performance of the self-healed composites, a composite sample that had already been tensile tested previously was healed by exposure to a heating plate for 3 min at 130 °C (Figure 6B). The self-healed composite was retested under tension and showed a similar mechanical behavior as in the initial measurement (Figure 6C). This observation is in agreement with previous findings for pristine vitrimers [13] and highlights the healing capabilities of the composite, which enables the extension of life cycles for such composite components. Future investigations regarding the composites' cycling capability are planned.

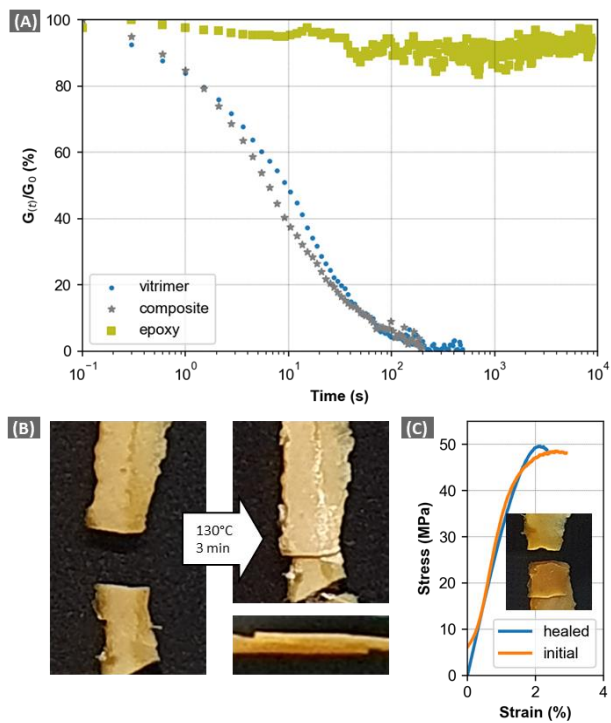

**Figure 6: (A) Normalized stress-relaxation curves for the vitrimer, an epoxy as thermosetting reference sample and a printed composite sample (monomer concentration 60 mg mL$^{-1}$, 24 h curing at 80 °C). (B) From left to right to bottom: Photograph of a fractured composite sample (60 mg mL$^{-1}$, 24 h at 80°C) previously tensile tested and two photographs from top and side view of the same sample after self-healing at 130 °C for 3 min. (C) Stress-strain curves of the sample presented in (C), which was initially tensile tested (initial), self-healed and retested under tension (healed).**

Next to fast and mechanically lossless self-healing capability, the composite additionally presents reshaping and shape memory capabilities (**Figure 7**). The initially planar and straight printed bar could be reshaped to a permanent spiral shape after 20 min exposure on a heating plate at 80 °C (Figure 7A-B). Due to its designed $T_g$ of 30 °C, the nacre-inspired composite spiral could be manually deformed just by the applied heat of the fingers (Figure 7C). The composite was shaped into a straight shape which was subsequently frozen temporarily when cooling down from 30 °C (Figure 7D). The stored spiral shape could be recovered by exposing the sample once again to the body temperature (Figure 7E) in less than 2.5 min. Fast recovery within ~10 s could be achieved on a heating plate at 60 °C. These reshaping and shape-memory capabilities allow for printing of planar nacre-inspired composites, thus preventing particle misalignment caused by e.g. particle radial shear alignment in the printing nozzle, and subsequent reshaping of the produced nacre-inspired composite to its final complex geometry. With such shape adaptation capabilities, the applicability of the printed components can be extended for multiple uses.

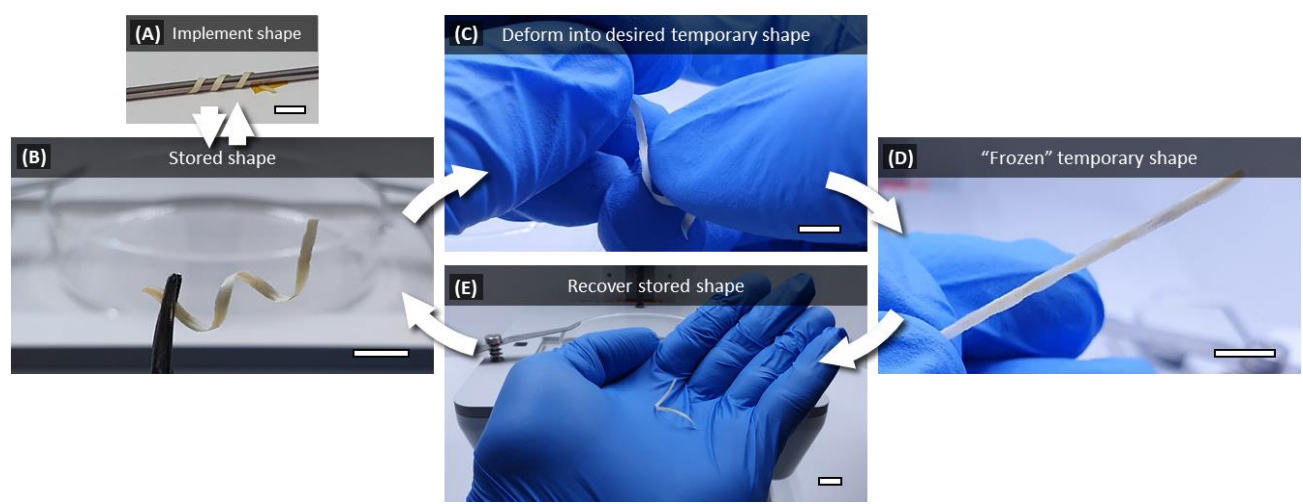

**Figure 7:** Photographs of (A) sample being reshaped on a heating plate at 80 °C for 20 min, (B) sample with new programmed shape, (C) sample being deformed into temporary shape above $T_g$ (=30 °C), (D) sample in temporary shape after cooling below $T_g$ and (E) sample recovering programmed shape upon exposure to human body temperature. Scale bars: 5mm.

## 4. Conclusion

In this work we presented a novel 4D-printing strategy for printing nacre-inspired composites. By developing the AMCA printing process of low viscous suspensions, the suspensions spread after extrusion, which induced shear alignment of platelets along a common plane. Thereby, planar nacre-inspired structures could be produced without the requirement of an additional external field for particle orientation[34]. Additionally, particle misalignment could be avoided that is omnipresent in printed structures of viscous or highly-laden suspensions, when the alignment of particles is governed by shear flow induced within the printing nozzle[37,38]. Furthermore, we demonstrated a novel *in situ* polymerization during the additive manufacturing process by taking advantage of solvent evaporation to tune the monomer concentration in the suspension and induce polycondensation reactions after extrusion. This approach enables printing thermosets via direct writing. Moreover, by surface funtionalizing the ceramic platelets, a covalently crosslinked matrix is formed through the composite material with covalent bonds being formed at the inorganic-organic interface. Thereby, no additional infiltration step is required after printing, since the whole composite is printed in a single step with a strong and stable interface between the organic matrix and inorganic platelets, as in natural nacre structures.

After curing, the printed nacre-inspired composites exhibited up to 3.3 and 26.7 times higher tensile strength and stiffness compared to the as cured pristine vitrimer, respectively, which was associated to successful load transfer mechanisms from the organic to the inorganic phase. Further toughening mechanisms such as crack deflection and platelet pullout could be observed in cured composites, which highlights the successful implementation of the nacre inspired structure.

By utilizing vinylogous vitrimers, not only we could print at ambient conditions without requiring an additional catalyst, but also the printed composite presented fast and mechanically lossless self-healing, reshaping, and shape memory capabilities. The 4D-printed nacre-inspired composite comprises a smart material system, with potential application on various fields, such as biomedicine[1,2] or aerospace[3–5]. Moreover, the presented 4D-printing strategy provides high design flexibility for nacre-inspired composites, which allows further optimization of the presented nacre-inspired composite and extension of the composite's functionalities in future studies.

**Acknowledgments**

The authors gratefully acknowledge the financial support from the Deutsche Forschungsgemeinschaft (DFG, German Research Foundation) – Projektnummer 192346071 – SFB 986, project C4, A2 and A6. The authors also gratefully acknowledge Magnus Küpker for his initial involvement in the project as well as Hans Jelitto for his valuable input regarding the mechanical characterization.

**References**

(1) Melocchi, A.; Uboldi, M.; Inverardi, N.; Briatico-Vangosa, F.; Baldi, F.; Pandini, S.; Scalet, G.; Auricchio, F.; Cerea, M.; Foppoli, A.; Maroni, A.; Zema, L.; Gazzaniga, A. Expandable drug delivery system for gastric retention based on shape memory polymers: Development via 4D printing and extrusion. *International journal of pharmaceutics* **2019**, *571*, 118700. DOI: 10.1016/j.ijpharm.2019.118700. Published Online: Sep. 14, 2019.
(2) Wei, H.; Zhang, Q.; Yao, Y.; Liu, L.; Liu, Y.; Leng, J. Direct-Write Fabrication of 4D Active Shape-Changing Structures Based on a Shape Memory Polymer and Its Nanocomposite. *ACS applied materials & interfaces* **2017**, *9* (1), 876–883. DOI: 10.1021/acsami.6b12824. Published Online: Dec. 20, 2016.

(3) Mat Yazik, M. H.; Sultan, M.T.H. Shape memory polymer and its composites as morphing materials. In *Failure Analysis in Biocomposites, Fibre-Reinforced Composites and Hybrid Composites*; Elsevier, 2019; pp 181–198. DOI: 10.1016/B978-0-08-102293-1.00009-7.

(4) Thill, C.; Etches, J.; Bond, I.; Potter, K.; Weaver, P. Morphing skins. *Aeronaut. j.* **2008**, *112* (1129), 117–139. DOI: 10.1017/S0001924000002062.

(5) Santo, L.; Quadrini, F.; Accettura, A.; Villadei, W. Shape Memory Composites for Self-deployable Structures in Aerospace Applications. *Procedia Engineering* **2014**, *88*, 42–47. DOI: 10.1016/j.proeng.2014.11.124.

(6) Tibbits, S. 4D Printing: Multi-Material Shape Change. *Archit Design* **2014**, *84* (1), 116–121. DOI: 10.1002/ad.1710.

(7) Khalid, M. Y.; Arif, Z. U.; Ahmed, W. 4D Printing: Technological and Manufacturing Renaissance. *Macro Materials & Eng* **2022**, *307* (8), 2200003. DOI: 10.1002/mame.202200003.

(8) Chen, X.; Liu, X.; Ouyang, M.; Chen, J.; Taiwo, O.; Xia, Y.; Childs, P. R. N.; Brandon, N. P.; Wu, B. Multi-metal 4D printing with a desktop electrochemical 3D printer. *Scientific reports* **2019**, *9* (1), 3973. DOI: 10.1038/s41598-019-40774-5. Published Online: Mar. 8, 2019.

(9) Lai, A.; Du, Z.; Gan, C. L.; Schuh, C. A. Shape memory and superelastic ceramics at small scales. *Science (New York, N.Y.)* **2013**, *341* (6153), 1505–1508. DOI: 10.1126/science.1239745.

(10) Cui, C.; Le An; Zhang, Z.; Ji, M.; Chen, K.; Yang, Y.; Su, Q.; Wang, F.; Cheng, Y.; Zhang, Y. Reconfigurable 4D Printing of Reprocessable and Mechanically Strong Polythiourethane Covalent Adaptable Networks. *Adv Funct Materials* **2022**, *32* (29), 2203720. DOI: 10.1002/adfm.202203720.

(11) Montarnal, D.; Capelot, M.; Tournilhac, F.; Leibler, L. Silica-like malleable materials from permanent organic networks. *Science (New York, N.Y.)* **2011**, *334* (6058), 965–968. DOI: 10.1126/science.1212648.

(12) Haida, P.; Signorato, G.; Abetz, V. Blended vinylogous urethane/urea vitrimers derived from aromatic alcohols. *Polym. Chem.* **2022**, *13* (7), 946–958. DOI: 10.1039/d1py01237a.

(13) Denissen, W.; Rivero, G.; Nicolaÿ, R.; Leibler, L.; Winne, J. M.; Du Prez, F. E. Vinylogous Urethane Vitrimers. *Adv Funct Materials* **2015**, *25* (16), 2451–2457. DOI: 10.1002/adfm.201404553.

(14) Haida, P.; Abetz, V. Acid-Mediated Autocatalysis in Vinylogous Urethane Vitrimers. *Macromolecular rapid communications* **2020**, *41* (16), e2000273. DOI: 10.1002/marc.202000273. Published Online: Jul. 30, 2020.

(15) Wegst, U. G. K.; Bai, H.; Saiz, E.; Tomsia, A. P.; Ritchie, R. O. Bioinspired structural materials. *Nature materials* **2015**, *14* (1), 23–36. DOI: 10.1038/NMAT4089. Published Online: Oct. 26, 2014.

(16) Bonderer, L. J.; Feldman, K.; Gauckler, L. J. Platelet-reinforced polymer matrix composites by combined gel-casting and hot-pressing. Part I: Polypropylene matrix composites. *Composites Science and Technology* **2010**, *70* (13), 1958–1965. DOI: 10.1016/j.compscitech.2010.07.014.

(17) Bonderer, L. J.; Studart, A. R.; Gauckler, L. J. Bioinspired design and assembly of platelet reinforced polymer films. *Science (New York, N.Y.)* **2008**, *319* (5866), 1069–1073. DOI: 10.1126/science.1148726.

(18) Bouville, F.; Maire, E.; Meille, S.; van de Moortèle, B.; Stevenson, A. J.; Deville, S. Strong, tough and stiff bioinspired ceramics from brittle constituents. *Nature materials* **2014**, *13* (5), 508–514. DOI: 10.1038/nmat3915. Published Online: Mar. 23, 2014.

(19) Cheng, Q.; Wu, M.; Li, M.; Jiang, L.; Tang, Z. Ultratough Artificial Nacre Based on Conjugated Cross-linked Graphene Oxide. *Angew. Chem.* **2013**, *125* (13), 3838–3843. DOI: 10.1002/ange.201210166.

(20) Erb, R. M.; Libanori, R.; Rothfuchs, N.; Studart, A. R. Composites reinforced in three dimensions by using low magnetic fields. *Science (New York, N.Y.)* **2012**, *335* (6065), 199–204. DOI: 10.1126/science.1210822.

(21) Gao, H.-L.; Chen, S.-M.; Mao, L.-B.; Song, Z.-Q.; Yao, H.-B.; Cölfen, H.; Luo, X.-S.; Zhang, F.; Pan, Z.; Meng, Y.-F.; Ni, Y.; Yu, S.-H. Mass production of bulk artificial nacre with excellent mechanical properties. *Nature communications* **2017**, *8* (1), 287. DOI: 10.1038/s41467-017-00392-z. Published Online: Aug. 18, 2017.

(22) Hunger, P. M.; Donius, A. E.; Wegst, U. G. K. Platelets self-assemble into porous nacre during freeze casting. *Journal of the mechanical behavior of biomedical materials* **2013**, *19*, 87–93. DOI: 10.1016/j.jmbbm.2012.10.013. Published Online: Nov. 3, 2012.

(23) Li, Y.-Q.; Yu, T.; Yang, T.-Y.; Zheng, L.-X.; Liao, K. Bio-inspired nacre-like composite films based on graphene with superior mechanical, electrical, and biocompatible properties. *Advanced materials (Deerfield Beach, Fla.)* **2012**, *24* (25), 3426–3431. DOI: 10.1002/adma.201200452.

(24) Lossada, F.; Jiao, D.; Hoenders, D.; Walther, A. Recyclable and Light-Adaptive Vitrimer-Based Nacre-Mimetic Nanocomposites. *ACS nano* **2021**, *15* (3), 5043–5055. DOI: 10.1021/acsnano.0c10001. Published Online: Feb. 25, 2021.

(25) Mirkhalaf, M.; Barthelat, F. Nacre-like materials using a simple doctor blading technique: Fabrication, testing and modeling. *Journal of the mechanical behavior of biomedical materials* **2016**, *56*, 23–33. DOI: 10.1016/j.jmbbm.2015.11.010. Published Online: Dec. 1, 2015.

(26) Munch, E.; Launey, M. E.; Alsem, D. H.; Saiz, E.; Tomsia, A. P.; Ritchie, R. O. Tough, bio-inspired hybrid materials. *Science (New York, N.Y.)* **2008**, *322* (5907), 1516–1520. DOI: 10.1126/science.1164865.

(27) Oner Ekiz, O.; Dericioglu, A. F.; Kakisawa, H. An efficient hybrid conventional method to fabricate nacre-like bulk nano-laminar composites. *Materials Science and Engineering: C* **2009**, *29* (6), 2050–2054. DOI: 10.1016/j.msec.2009.04.001.

(28) Podsiadlo, P.; Kaushik, A. K.; Arruda, E. M.; Waas, A. M.; Shim, B. S.; Xu, J.; Nandivada, H.; Pumplin, B. G.; Lahann, J.; Ramamoorthy, A.; Kotov, N. A. Ultrastrong and stiff layered polymer nanocomposites. *Science (New York, N.Y.)* **2007**, *318* (5847), 80–83. DOI: 10.1126/science.1143176.

(29) Wang, J.; Song, T.; Chen, H.; Ming, W.; Cheng, Z.; Liu, J.; Liang, B.; Wang, Y.; Wang, G. Bioinspired High-Strength Montmorillonite-Alginate Hybrid Film: The Effect of Different Divalent Metal Cation Crosslinking. *Polymers* **2022**, *14* (12). DOI: 10.3390/polym14122433. Published Online: Jun. 16, 2022.

(30) Wu, H.; Li, J.; Zhang, W.; Chen, T.; Liu, F.; Han, E.-H. Supramolecular engineering of nacre-inspired bio-based nanocomposite coatings with exceptional ductility and high-efficient self-repair ability. *Chemical Engineering Journal* **2022**, *437*, 135405. DOI: 10.1016/j.cej.2022.135405.

(31) Yu, Y.; Kong, K.; Tang, R.; Liu, Z. A Bioinspired Ultratough Composite Produced by Integration of Inorganic Ionic Oligomers within Polymer Networks. *ACS nano* **2022**, *16* (5), 7926–7936. DOI: 10.1021/acsnano.2c00663. Published Online: Apr. 28, 2022.

(32) Zhao, C.; Zhang, P.; Zhou, J.; Qi, S.; Yamauchi, Y.; Shi, R.; Fang, R.; Ishida, Y.; Wang, S.; Tomsia, A. P.; Liu, M.; Jiang, L. Layered nanocomposites by shear-flow-induced alignment of nanosheets. *Nature* **2020**, *580* (7802), 210–215. DOI: 10.1038/s41586-020-2161-8. Published Online: Apr. 8, 2020.

(33) Behr, S.; Vainio, U.; Müller, M.; Schreyer, A.; Schneider, G. A. Large-scale parallel alignment of platelet-shaped particles through gravitational sedimentation. *Scientific reports* **2015**, *5*, 9984. DOI: 10.1038/srep09984. Published Online: May. 18, 2015.

(34) Zhou, X.; Ren, L.; Liu, Q.; Song, Z.; Wu, Q.; He, Y.; Li, B.; Ren, L. Advances in Field-Assisted 3D Printing of Bio-Inspired Composites: From Bioprototyping to Manufacturing. *Macromolecular bioscience* **2022**, *22* (3), e2100332. DOI: 10.1002/mabi.202100332. Published Online: Dec. 16, 2021.

(35) Martin, J. J.; Fiore, B. E.; Erb, R. M. Designing bioinspired composite reinforcement architectures via 3D magnetic printing. *Nature communications* **2015**, *6*, 8641. DOI: 10.1038/ncomms9641. Published Online: Oct. 23, 2015.

(36) Feilden, E.; Ferraro, C.; Zhang, Q.; García-Tuñón, E.; D'Elia, E.; Giuliani, F.; Vandeperre, L.; Saiz, E. 3D Printing Bioinspired Ceramic Composites. *Scientific reports* **2017**, *7* (1), 13759. DOI: 10.1038/s41598-017-14236-9. Published Online: Oct. 23, 2017.

(37) Walton, R. L.; Brova, M. J.; Watson, B. H.; Kupp, E. R.; Fanton, M. A.; Meyer, R. J.; Messing, G. L. Direct writing of textured ceramics using anisotropic nozzles. *Journal of the European Ceramic Society* **2021**, *41* (3), 1945–1953. DOI: 10.1016/j.jeurceramsoc.2020.10.021.

(38) Li, T.; Liu, Q.; Qi, H.; Zhai, W. Prestrain Programmable 4D Printing of Nanoceramic Composites with Bioinspired Microstructure. *Small (Weinheim an der Bergstrasse, Germany)* **2022**, *18* (47), e2204032. DOI: 10.1002/smll.202204032. Published Online: Sep. 30, 2022.

(39) Ding, F.; Liu, J.; Zeng, S.; Xia, Y.; Wells, K. M.; Nieh, M.-P.; Sun, L. Biomimetic nanocoatings with exceptional mechanical, barrier, and flame-retardant properties from large-scale one-step coassembly. *Science advances* **2017**, *3* (7), e1701212. DOI: 10.1126/sciadv.1701212. Published Online: Jul. 19, 2017.

(40) Nayak, N.; Huertas, R.; Crespo, J. G.; Portugal, C. A.M. Surface modification of alumina monolithic columns with 3-aminopropyltetraethoxysilane (APTES) for protein attachment. *Separation and Purification Technology* **2019**, *229*, 115674. DOI: 10.1016/j.seppur.2019.115674.

(41) Yamaguchi, A.; Uejo, F.; Yoda, T.; Uchida, T.; Tanamura, Y.; Yamashita, T.; Teramae, N. Self-assembly of a silica-surfactant nanocomposite in a porous alumina membrane. *Nature materials* **2004**, *3* (5), 337–341. DOI: 10.1038/nmat1107. Published Online: Apr. 11, 2004.

(42) Xu, H.; Wang, H.; Zhang, Y.; Wu, J. Vinylogous Urethane Based Epoxy Vitrimers with Closed-Loop and Multiple Recycling Routes. *Ind. Eng. Chem. Res.* **2022**, *61* (48), 17524–17533. DOI: 10.1021/acs.iecr.2c03393.

(43) Zhang, Y.; Evans, J. R. G. Morphologies developed by the drying of droplets containing dispersed and aggregated layered double hydroxide platelets. *Journal of colloid and interface science* **2013**, *395*, 11–17. DOI: 10.1016/j.jcis.2012.09.089. Published Online: Jan. 8, 2013.

(44) Wang, X.; Lin, K. S. K.; Chan, J. C. C.; Cheng, S. Direct synthesis and catalytic applications of ordered large pore aminopropyl-functionalized SBA-15 mesoporous materials. *The journal of physical chemistry. B* **2005**, *109* (5), 1763–1769. DOI: 10.1021/jp045798d.

(45) Chen, T.; Niu, M.; Xie, Y.; Wu, Z.; Liu, X.; Cai, L.; Zhuang, B. Modification of Ultra-Low Density Fiberboards by an Inorganic Film Formed by Si-Al Deposition and their Mechanical Properties. *BioResources* **2014**, *10* (1). DOI: 10.15376/biores.10.1.538-547.

(46) Ebrahimi-Gatkash, M.; Younesi, H.; Shahbazi, A.; Heidari, A. Amino-functionalized mesoporous MCM-41 silica as an efficient adsorbent for water treatment: batch and fixed-bed column adsorption of the nitrate anion. *Appl Water Sci* **2017**, *7* (4), 1887–1901. DOI: 10.1007/s13201-015-0364-1.

(47) Salili, S. M.; Ataie, A.; Barati, M. R.; Sadighi, Z. Characterization of mechano-thermally synthesized Curie temperature-adjusted La0.8Sr0.2MnO3 nanoparticles coated with (3-aminopropyl) triethoxysilane. *Materials Characterization* **2015**, *106*, 78–85. DOI: 10.1016/j.matchar.2015.05.025.

(48) Menon, M.; Decourcelle, S.; Ramousse, S.; Larsen, P. H. Stabilization of Ethanol-Based Alumina Suspensions. *J American Ceramic Society* **2006**, *89* (2), 457–464. DOI: 10.1111/j.1551-2916.2005.00744.x.

(49) Khimi, S. R.; Pickering, K. L. A new method to predict optimum cure time of rubber compound using dynamic mechanical analysis. *J. Appl. Polym. Sci.* **2014**, *131* (6), n/a-n/a. DOI: 10.1002/APP.40008.

(50) Lee, J. B.; Derome, D.; Guyer, R.; Carmeliet, J. Modeling the Maximum Spreading of Liquid Droplets Impacting Wetting and Nonwetting Surfaces. *Langmuir : the ACS journal of surfaces and colloids* **2016**, *32* (5), 1299–1308. DOI: 10.1021/acs.langmuir.5b04557. Published Online: Jan. 25, 2016.

(51) Rocha, B. C.; Paul, S.; Vashisth, H. Role of Entropy in Colloidal Self-Assembly. *Entropy (Basel, Switzerland)* **2020**, *22* (8). DOI: 10.3390/e22080877. Published Online: Aug. 10, 2020.

(52) Santhiya, D.; Subramanian, S.; Natarajan, K. A.; Malghan, S. G. Surface chemical studies on alumina suspensions using ammonium poly(methacrylate).

Supporting Information

## 4D-Printing of Smart, Nacre-Inspired, Organic-Ceramic Composites

*Benedikt F. Winhard, Philipp Haida, Alexander Plunkett, Julian Katz, Berta Domènech, Volker Abetz, Kaline P. Furlan, Gerold A. Schneider*

### 1. Alumina platelet functionalization

In order to estimate the required amount of silane ligands to functionalize the total surface area of the used alumina platelets, a platelet cylindrical shape was assumed with the surface $A = 2\pi r(r + h)$ with $r = d_{50}/2 = 1.0$ µm and $h = 0.04$ µm being the radius and height of the platelets, respectively. Consequently, the specific surface area per gram of particles was obtained by using the density, $\rho$, and the volume, $V = \pi r^2 h$, with

$$\frac{A}{m_{Al_2O_3}} = \frac{A}{\rho V} = 2\frac{r+h}{\rho r h}$$

giving a specific surface area of 13 m² g⁻¹. Assuming a surface area of $a_{silane} = 0.27$ nm² for the silane head group, one surface equivalent of silane molecules per gram of platelets can be estimated using the molecular weight of the silane ligand, $M_{silane} = 221.37$ g mol⁻¹, and the Avogadro number, $N_A$, with

$$\frac{m_{silane}}{m_{Al_2O_3}} = \frac{A}{m_{Al_2O_3}} \frac{M_{silane}}{N_A a_{silane}}$$

giving 0.019 g$_{silane}$ g⁻¹$_{Al2O3}$ corresponding to 1 surface equivalent. Small errors due to the assumed silane head group area are insignificant considering the large scatter of platelet diameters and shapes.

## 2. Suspension's rheology

A shear rate ramp was performed for the suspension ($c$(Al$_2$O$_3$) = 40 mg mL$^{-1}$, $c$(monomers) = 60 mg mL$^{-1}$) as well as the solvent ethanol as reference. For the determination of the flow index $n$ the results for the viscosity of the suspension were fitted with the Ostwald-de Waele model (Figure S1A), for the yield stress of the suspension, the corresponding results for the shear stress were fitted with the Herschel-Bulkley model (Figure S1A). In contraposition to the reference ethanol, the suspension is low viscous, highly shear thinning with a flow index $n$ = 0.29 and possesses a small yield stress of $\tau_0$ = 18 mPa. The vertical dashed line in Figure S1A indicates the estimated average shear rate in the spreading suspension $\dot{\gamma}_S$ ~200 s$^{-1}$.

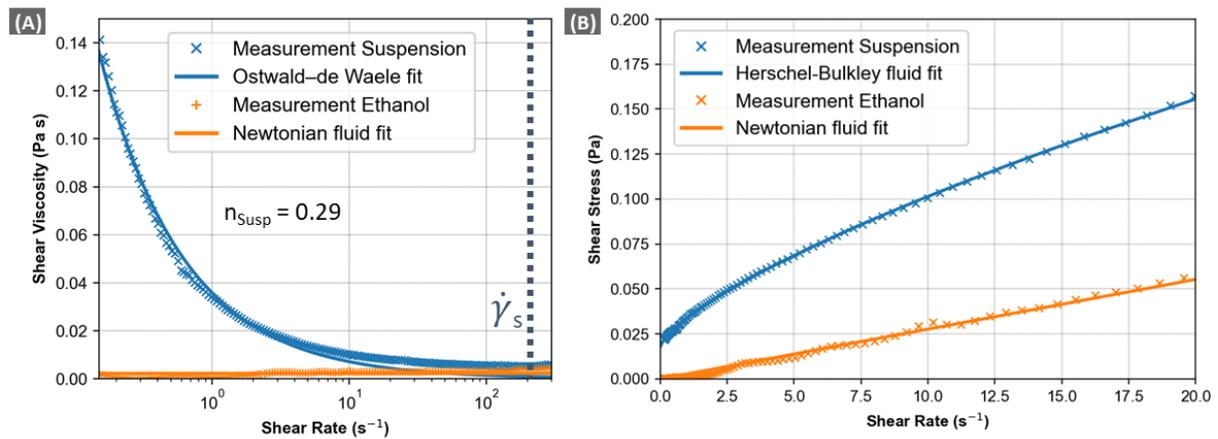

**Figure S1: Shear rate ramp for the printing suspension ($c$(Al$_2$O$_3$) = 40mg mL$^{-1}$, $c$(monomers) = 60mg mL$^{-1}$) as well as the pristine solvent ethanol as reference. (A) shear viscosity over shear rate: the measured data of the suspension was fitted with the Ostwald-de Waele model; the flow index $n$ determined by the fit is stated in the graph and the estimate for the shear rate of the spreading suspension $\dot{\gamma}_S$ is marked with a dashed line (B) shear stress over shear rate: the measured data of the suspension was fitted with the Herschel-Bulkley model. To fit the Newtonian behavior of ethanol in (A) and (B) a linear fit was used.**

## 3. Wetting of suspension on various substrates

To analyze the wetting of the suspension (40 mg mL$^{-1}$ functionalized platelets, total monomer concentration of 10 mg mL$^{-1}$) on different substrates that could be suitable for printing, single droplets (droplet volume ~0.3 µL) were deposited on the substrates: Glass, silanized glass, PMMA, and PVA showed total wetting near the contact lines. While the organic phase could migrate towards the contact line during total wetting, the platelets couldn't, thus remained in a confined space. On Parafilm no total wetting, but an equilibrium contact angle of 16.7 ±0.3 ° was determined. The reduced wetting on Parafilm allowed the platelets to migrate to the contact line, thus both phases could simultaneously spread over the entire wetted area.

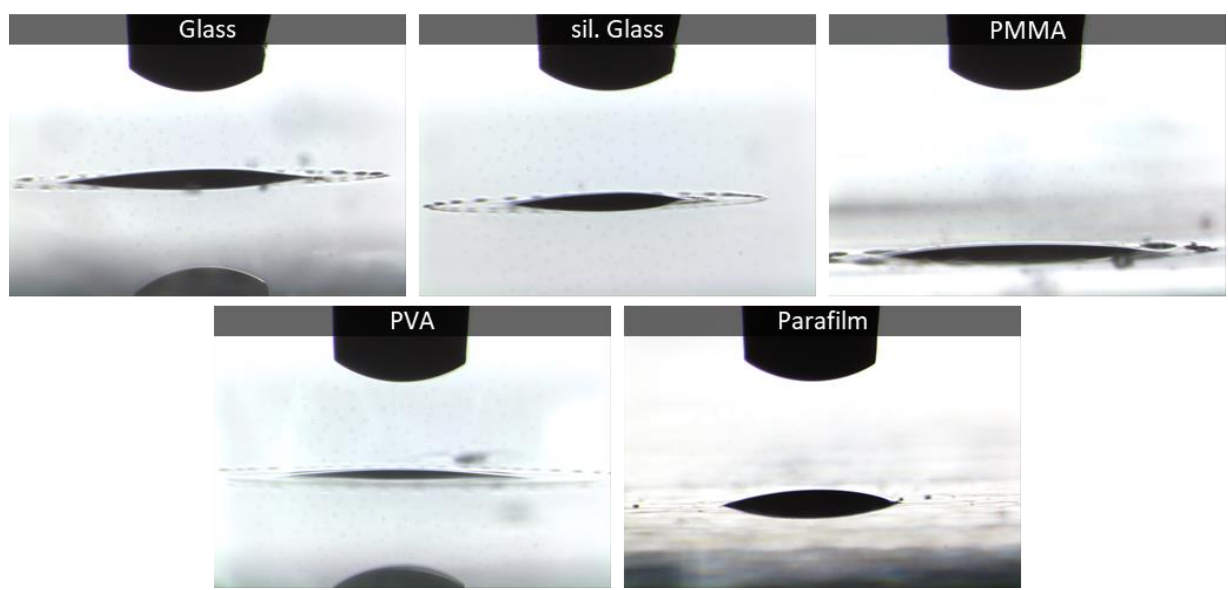

**Figure S2:** Cross sectional view of sessile droplets right after deposition on various substrates, when equilibrium contact angle was reached.

## 4. ATR-FTIR spectra for various curing stages of the printed composites

ATR-FTIR spectra of the composite in the uncured (0 h) as well as in various cured states (3 h, 12 h, 24 h) revealed the characteristic absorption peaks of the vitrimer, confirming the formation of the vitrimer network during printing (C=O: 1647 cm$^{-1}$; C=C: 1590 cm$^{-1}$). Additionally, a broad increase in absorption in the 860-600 cm$^{-1}$ range (Al-O bond[52]), proves the presence of alumina platelets in the printed composites.

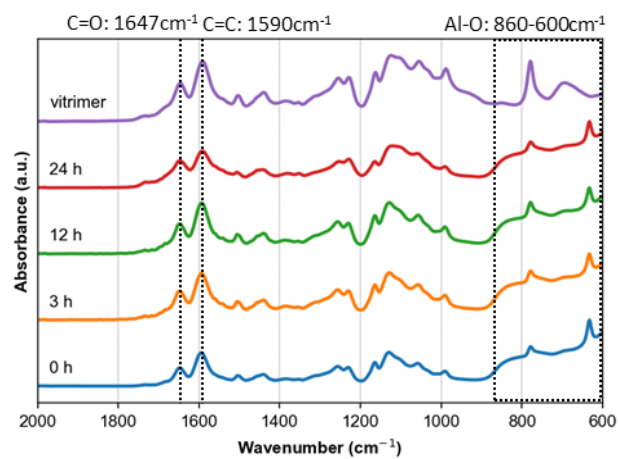

**Figure S3: ATR-FITR spectra for the vitrimer as well as the printed composite after different curing durations (0, 3, 12, 24 h)**

## 5. DSC curves of the vitrimer and the 24h-cured composite

Four temperature cycles were performed with each sample (vitrimer and 24 h-cured composite) under $N_2$: In the first two cycles, the samples were heated from -50 °C to 80 °C and cooled down to -50 °C. In the third and fourth cycle the samples were heated again from -50 °C but up to 150 °C. The heating and cooling rate was 10 K min$^{-1}$. The first cycle was used for priming, and only the last three cycles were considered for further analysis, hence we refer to the original second cycle as the first cycle. As can be seen in the DSC curves for the heating segments, both samples kept a *Tg* of 30 °C, after each cycle.

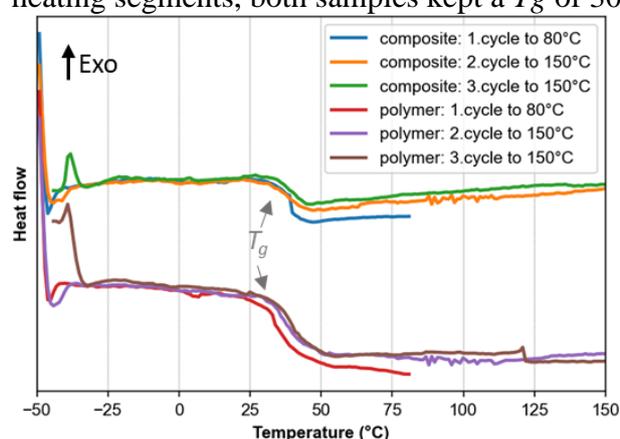

**Figure S4: Differential scanning calorimetrie measurements of the vitrimer and a printed 24 h-cured composite ($c = 60$ mg mL$^{-1}$).**

## 6. Temperature-dependent DMA spectra of the vitrimer and the 24 h-cured composite

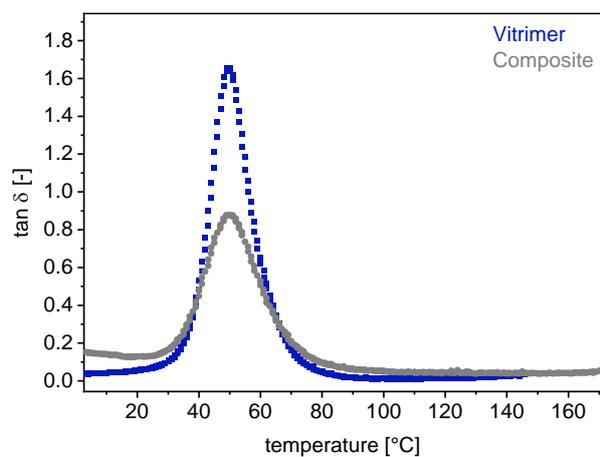

**Figure S5: Temperature-dependent DMA, confirming the formation of crosslinked vitrimers and composites by measuring the storage and loss moduli at 0-170 °C and plotting tan $\delta$ against the temperature ($\omega = 10$ rad s$^{-1}$, $\gamma = 0.1$ %).**